\documentclass[aps,prb,reprint]{revtex4-1}
\usepackage[dvipdfm]{graphicx}
\usepackage[dvipdfm]{rotating}
\usepackage{subfigure}
\usepackage{amsmath}
\usepackage{amsfonts}
\usepackage{amssymb}

\begin{document}

\title{Diffusion Mechanisms in Lithium Disilicate Melt by Molecular Dynamics Simulation}
\author{Luis G. V. Gon\c{c}alves}
\email{vieira@df.ufscar.br}
\affiliation{Departamento de F\'{\i}sica, Universidade Federal de S\~{a}o Carlos,
13.565-905, S\~{a}o Carlos, S\~{a}o Paulo, Brazil}
\author{Jos\'e P. Rino}
\affiliation{Departamento de F\'{\i}sica, Universidade Federal de S\~{a}o Carlos,
13.565-905, S\~{a}o Carlos, S\~{a}o Paulo, Brazil}

\begin{abstract}

In this work we study the diffusion mechanisms in lithium disilicate melt using molecular dynamics simulation, which has an edge over other simulation methods because it can track down actual atomic rearrangements in materials once a realistic interaction potential is applied. Our simulation results of diffusion coefficients show an excellent agreement with experiments. We also demonstrate that our system obeys the famous Stokes-Einstein relation at least down to 1400 K, while a decoupling between relaxation and viscosity takes place at a higher temperature. Additionally, an analysis on the dynamical behavior of slow-diffusing atoms reveals explicitly the presence of dynamical heterogeneities. 

\end{abstract}


\keywords{diffusion, activated jumps, undercooled liquids, relaxation, silicate glasses }

\maketitle

\section{Introduction}

The dynamics of complex systems has challenged scientists for many years since the first attempts on explaining the glass transition\cite{Debenedetti1997,Stillinger1995,Kob1997,Debenedetti2001}. Glasses are among the most important materials of technological interest and thorough understanding of their synthesis is crucial. It is a well-known fact that diffusion controls processes like nucleation, growth, and electrical conductivity in liquids. Many theories rely upon this assumption. Therefore, information about mobility of atoms is of great importance. In the case of glass-ceramics\cite{Zanotto2010}, for example, one has to control nucleation and growth in the supercooled liquid to obtain the desired outcome. 

Despite of its importance, however, diffusion coefficients are not always easily accessible through experiments. One largely employed method is to use tracers, which are atoms that can be monitored while they move within the material analyzed. The problem is that isotopes that can be used as tracers are very expensive. A cheaper alternative is to employ an effective diffusion coefficient obtained via the Stokes-Eintein or the Eyring relation. They state that diffusion divided by temperature is inversely proportional to the viscosity of the liquid. Both relations differ only by a multiplicative factor. Determining viscosity is a much easier task and data for several materials are abundant.

There is still an ongoing debate about the range of validity of the Eyring relation. The relation is suitable for ordinary liquids above the melting temperature. As we approach the glass transition region, dynamical heterogeneities in the liquid emerge and a novel diffusion regime takes place\cite{Sastry1998}. Molecular dynamics have proven to be very useful to describe atomic processes on a nanometric scale. The following questions will be discussed in this work: 1- How can molecular dynamics assess the range of validity of the Eyring relation? 2- How do atoms rearrange in the system near the glass transition? 3- Are dynamical heterogeneities \emph{directly} responsible for the breakdown of the Eyring relation?.

According to recent studies, the breakdown of the Eyring relation depends mostly on the material's fragility. Studies on silica\cite{Nascimento2006}, a strong glass former, reveal that the Eyring relation holds all the way down to the calorimetric glass transition temperature $T_g$. On the other hand, it is shown for fragile liquids\cite{Nascimento2010} a breakdown near 1.1$T_g$. Simulations on model glass formers such as the binary Lennard-Jones liquid also point to a breakdown. This breakdown also occurs far from $T_g$ according to a recent study on supercooled water confined in nanopores\cite{Chen2006}.

In this work, classical molecular dynamics simulation will be used to model lithium disilicate melt. It is an archetypal fragile oxide liquid for which transport data are numerous. Simulation results of diffusion and relaxation will be compared to the experimental data available. Also, a numerical experiment devised to track molecular rearrengements was performed. This work aims to shed light on the aforementioned questions and to give alternative interpretations about computer simulation results.

\section{Computational details}
\label{sec:details}

Classical molecular-dynamics (MD) simulation is the main computational method employed in this work. It is based on integrating the coupled newtonian equations of motion for a set of interacting particles. The particle interaction used to model lithium disilicate is the Buckingham potential, with parameters developed by Habasaki \emph{et al.}\cite{Habasaki95}. The potential has the form:

\begin{equation}
V_{ij}(r)=\frac{q_iq_je^2}{r}-\frac{C_{ij}}{r}+A_{ij}\exp(-B_{ij}r)
\end{equation}
where $r$ is the interparticle distance, $i$ and $j$ are the species, $q$ is the effective charge for coulombic interactions and $A$, $B$ and $C$ are fitting parameters. This interatomic potential has proven to describe structural properties in the melt and the amorphous phases in good agreement with neutron scattering experiments\cite{}. Further details on the potential can be found in Ref. \onlinecite{Banhatti01}.

All computer simulations were performed with the LAMMPS package, which is developed by S. Plimpton and co-workers\cite{Plimpton95}. A time step of 1.0 fs was used in the entire work. Periodic boundary conditions were applied in all directions to simulate the bulk material. The long-range forces due to electrostatic interactions were properly handled using the Particle-Particle Particle-Mesh (PPPM) technique\cite{Hockney89}, with a real space cutoff of 10 \AA\: and energy accuracy of 1$\times$10$^\text{-4}$.

A system with 3888 atoms was setup initially in the orthorhombic structure for Li$_2$Si$_2$O$_5$\cite{Liebau61,Du06} at 300 K. To obtain liquid configurations, we heated the crystalline system up to 3000 K in a constant-pressure ensemble at zero pressure. Finally, the hot liquid was cooled down at a desired temperature and allowed to thermalize in a constant-volume ensemble. Thermostatting was done using the Nos\'e-Hoover chain method\cite{Martyna92}. During all stages, the simulation box was kept orthogonal in order to properly simulate the liquid state.

The liquid properties were calculated after adequate thermalization. The self-diffusion coefficient $D_i$ for the $i$-element was calculated via the Einstein relation\cite{Allen89}

\begin{equation}
 D_i=\lim_{t\rightarrow\infty}\frac{1}{6t}\left\langle r_i^2(t)\right\rangle
\end{equation}
where $\left\langle r_i^2(t)\right\rangle$ is the mean-square displacement at a time $t$. The structural relaxation time $\tau_\alpha$ was calculated from the intermediate scattering function\cite{Heuer08}

\begin{equation}
 F_s(\vec{k},t)=\left\langle\cos[\vec{k}\cdot(\vec{r}(t)-\vec{r}(0))]\right\rangle
\end{equation}
and via fitting of

\begin{equation}\label{eq:relax}
 F_s(\vec{k},t)=\exp\left(-\frac{t}{\tau_\alpha}\right)^\beta
\end{equation}
where $\vec{k}$ is the reciprocal vector of the first sharp peak of the static structure factor and $\vec{r}(t)$ is the position of a particle at time $t$. The structural relaxation time is proportional to the shear viscosity $\eta$ via

\begin{equation}\label{eq:Gtau}
\eta(T)=G_\infty\tau_\alpha(T)
\end{equation}
where $G_\infty$ is the instantaneous shear modulus and $T$ is the temperature. $G_\infty$ is weakly dependent of the temperature\cite{Heuer08}. Hence, $\tau_\alpha$ and $\eta$ have a very similar temperature dependence.

The structural relaxation is a measure of how long the liquid structure rearranges globally due to viscous flow. In the case of all alkali-silicates, those rearrangements are due to the mobility of the network-formers, i.e. SiO$_4$ tetrahedra. For that reason, $\tau_\alpha$ was calculated for Si atoms only. 

\section{Results and analyses}

The primary goal of this work is to study the relation between diffusion and viscosity using the Eyring equation (see Sec.\ref{sec:eyring}) and analyze its range of validity. To do this, we had to calculate $D$ and $\tau_\alpha$ as a function of the temperature with the highest degree of undercooling as possible. A great care was taken in those calculations. $D$ in the diffusive regime is properly calculated if, and only if, the mean-square displacement is linear with $t$ for a prolonged period of time. Calculating $D$ on a sub-diffusive regime ($\left\langle r^2(t)\right\rangle\propto t^\gamma,\gamma<$1) only gives us an upper limit for $D$. Also, to obtain a good fit for $\tau_\alpha$, $F_s(t)$ should be at least below 1/$e$\cite{Heuer08}. We have calculated $D$ and $\tau_\alpha$ for Si atoms down to 1200 K, to which it was necessary a simulation run of 40 ns.

The experimental value of the liquidus temperature for lithium disilicate is 1360 K. The orthorhombic-liquid coexistence temperature\cite{Hakkinen1992} calculated in this work is equal to 1450$\pm$30 K, in good agreement with experiment. Hence, MD simulation, in this case, could only achieve a small degree of undercooling to calculate relaxation and slowly diffusing atoms. Nevertheless, effects of undercooling such as non-exponential relaxation were observed in our work. We have not detected any signs of crystallization during simulations, even at the highest degree of supercooling.

\subsection{Diffusion}
\label{sec:diffusion}

\begin{figure}[htb!]
 \begin{center}
 \includegraphics[width=6.0cm,angle=270]{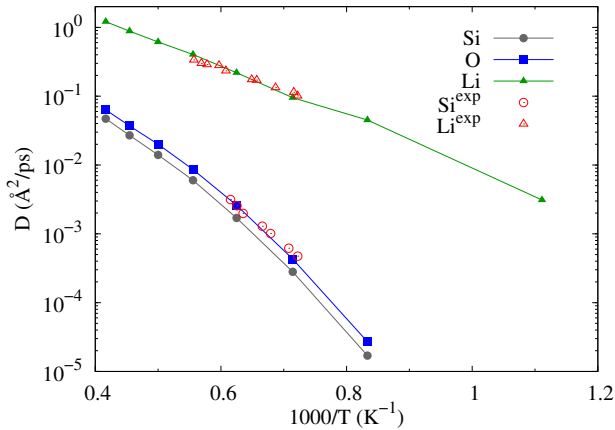}
 \caption{Self-diffusion coefficients as a function of the reciprocal temperature. Open symbols are experimental data for Li (Ref. \onlinecite{Bockris52}) and Si (Ref. \onlinecite{Kawakami78}).}
 \label{diff}
 \end{center}
\end{figure}

The temperature dependence of the self-diffusion for all three elements in lithium disilicate is plotted in Fig. \ref{diff}. Along with MD data (filled symbols) are experimental data (open symbols) available for comparison. Si diffusion was measured experimentally by electrochemical determination of cations interdiffusivity\cite{Kawakami78}. Conductivity data at high temperatures\cite{Bockris52} were used to calculate Li diffusion via the Nernst-Einstein relation\cite{Souquet10}. An excellent agreement is found in Li diffusion between MD and experimental values. For the silicon diffusion, absolute values agree satisfactorily. The temperature dependence also shows excellent agreement with experiment. Overall, Fig. \ref{diff} displays the typical behavior for an alkali silicate melt. The system presents two distinct diffusion mechanisms\cite{Avramov2009}: a) slow Si and O atoms, which move as an SiO$_\text{4}$ unit (i.e. the network formers), and b) the light, fast moving Li atoms, which permeate the network and act as the network modifiers. It is believed that the former mechanism is related to viscous flow over a wide temperature range accessible to experiments. This topic will be discussed on the next sections.

\subsection{Relaxation and the Eyring relation}
\label{sec:eyring}

\begin{figure}[htb!]
 \begin{center}
 \includegraphics[width=6.0cm,angle=270]{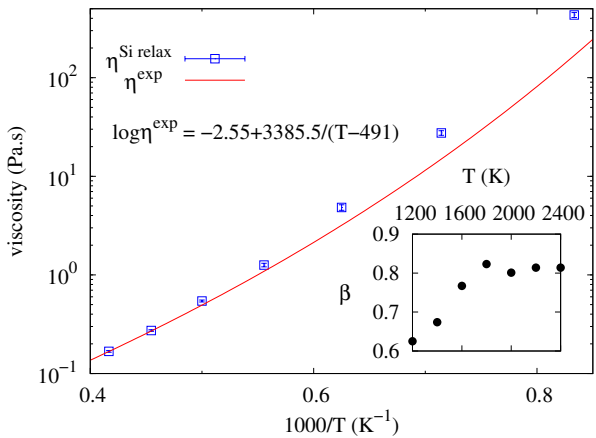}
 \caption{Temperature dependence of relaxation-defined and experimental viscosities.}
 \label{relax}
 \end{center}
\end{figure}

The intermediate scattering function was obtained and $\tau_\alpha$ was calculated from the fitting of Eq. \ref{eq:relax}. This value of $\tau_\alpha$ in the equation \ref{eq:Gtau} results in an effective viscosity $\eta^{relax}$, which allows direct comparison with experimental viscosity data. Using $\tau_\alpha$ and the experimental viscosity value at 2200 K in Eq. \ref{eq:Gtau}, we obtained $G_\infty$ equal to 17.5 GPa. In Fig. \ref{relax}, $\eta^{relax}$ and a fit of the Vogel-Fulcher-Tammann function of experimental viscosity data\cite{Burgner2001} are plotted against temperature. One can observe that both properties are incompatible except at the high temperature regime $T>$2000 K. Below that regime, the relaxation-defined viscosity is always larger than the actual data.

The parameter $\beta$ was calculated via fitting of Eq.\ref{eq:relax} (inset of Fig.\ref{relax}). If $\beta=1.0$ the relaxation is exponential. It was shown that the system has non-exponential relaxation for all temperatures analyzed. At 2400K (1000K above the melting temperature), the exponent $\beta$ is close to 0.8 and decreases with further cooling. It is generally accepted that non-exponential relaxation is due to spatial heterogeneities found in the liquid upon supercooling. Note that in this silicate model, no undercooling was necessary to observe such behavior. This is an indication that the non-Arrhenian behavior of the viscosity found in fragile liquids far above $T_g$ is related to spatial heterogeneities.

\begin{figure}[htb!]
 \begin{center}
 \includegraphics[width=6.0cm,angle=270]{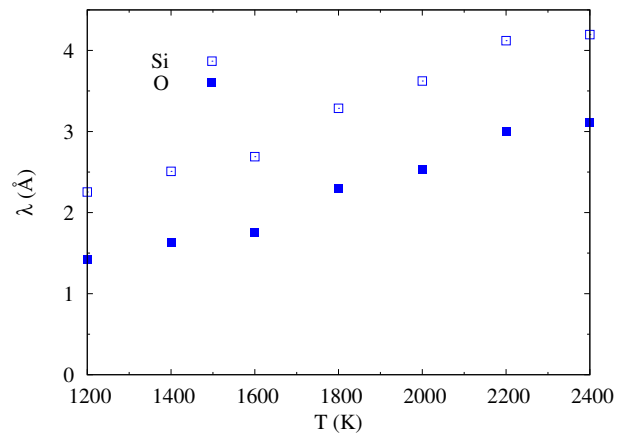}
 \caption{Temperature dependence of the jump distance obtained using molecular-dynamics data.}
 \label{eyring}
 \end{center}
\end{figure}

It is shown in Fig. \ref{eyring} the temperature dependence of the characteristic structural length $\lambda$ calculated via the Eyring relation\cite{Gutzow95} 

\begin{equation}\label{eq:eyring}
D=\frac{k_BT}{\lambda\eta^\text{relax}}
\end{equation}
where $\lambda$ is a constant. Thus, the Eyring relation does not hold at any temperature using diffusion and relaxation calculated via MD. This observation had already been reported by Bordat \emph{et al.}\cite{Bordat2003} for a binary Lennard-Jones model, and Saika-Voivod \emph{et al.}\cite{Saika-Voivod2009} for a high-pressure \emph{fragile} silica model. On the other hand, a simulation study on a metallic glass model\cite{Das2008} had shown that decoupling between diffusion and relaxation could only occur if the system were subject to deep undercooling. In our scenario, the degree of undercooling clearly does not play a role in this kind of breakdown.

The interpretation of $\lambda$ comes from the original concept of Eq.\ref{eq:eyring} as a hydrodynamic model and it is equal to $6\pi r_0$, where $r_0$ is the apparent hydrodynamic radius. However, as the temperature decreases, particles have their mobility restricted to cages of decreasing volume and diffusion becomes mediated by atomic jumps. The association of the characteristic length $\lambda$ and the size of the cage is still on debate and we will restrict the discussion of jump diffusion in the next section.

\begin{figure}[htb!]
 \begin{center}
 \includegraphics[width=6.0cm,angle=270]{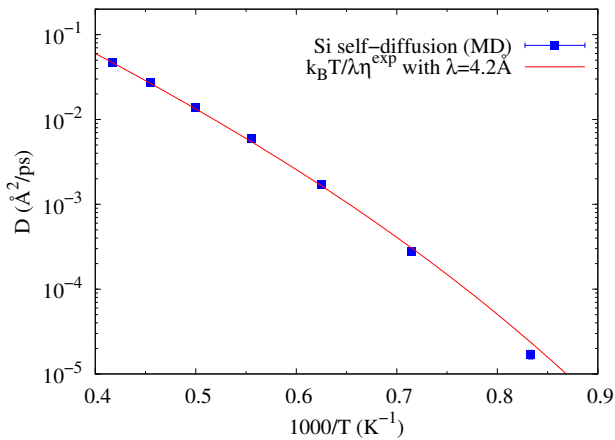}
 \caption{The calculated Si-diffusion coefficient and the Eyring relation as a function of temperature. A single jump distance $\lambda=$4.2 \AA \: was sufficient to render the Eyring relation valid all the way down to 1200 K. Experimentally measured viscosity data were used in this plot.}
 \label{eyringExp}
 \end{center}
\end{figure}

If $\eta^\text{relax}$ is replaced by the experimental viscosity, the Eyring relation is valid all the way down to 1400 K. We applied a single $\lambda=$ 4.2 \AA to the experimental data to allow comparison with diffusion calculated directly via MD (Fig. \ref{eyringExp}). Indeed, there are several experimental studies \cite{Nascimento2006,Reinsch2008,Andraca2008} showing that the Eyring relation is valid over a wide range of temperatures. Some studies on oxide glasses\cite{Ediger2008,Nascimento2010} suggest a breakdown of the Eyring relation near the glass-transition temperature. Unfortunately, straightforward MD simulation is not capable to calculate diffusion at such low temperatures for silicate glasses. Still, it is worth mentioning that calculating diffusion coefficients at temperatures as low as 1200 K for this system can be problematic due to the prolonged relaxations found in these conditions. Nevertheless, the trends observed in Fig. \ref{eyringExp} are remarkable given the extreme conditions of the simulation.

\subsection{Description of jump mediated diffusion}
\label{sec:jump}

\begin{figure}[htb!]
 \begin{center}
 \includegraphics[width=6.0cm,angle=270]{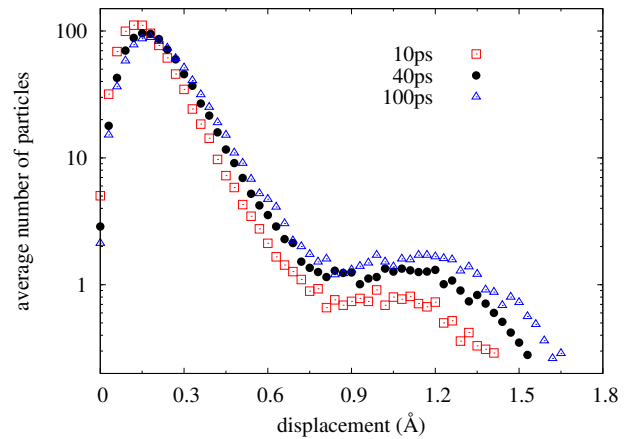}
 \caption{Distribution of particle displacements between two adjacent configurations. Time intervals of 10 (squares), 40 (circles) and 100ps (triangles) were used.}
 \label{dist}
 \end{center}
\end{figure}

Another important question on the dynamics of fragile glass formers is about the way molecular rearrangements occur. Down to 1200K, atomic motion becomes rather sluggish and straightforward visual inspection is inefficient. Based on the technique of A. Voter\cite{Voter1998,Voter2002} devised to identify infrequent events in many-body systems, we setup the following procedure: 1- Starting from a equilibrated configuration at 1200K, an MD simulation is carried out in the NVT ensemble; 2- A series of atomic configurations are recorded, being each sample separated by some fixed time interval; 3- These configurations are minimized using a standard conjugate-gradient algorithm; 4- Si atoms displacements are calculated between two adjacent minimized configurations. The distribution of the displacements calculated for all minimized configurations are plotted in Fig.\ref{dist}.

In agreement with several studies, this model for lithium disilicate also presented dynamical heterogeneities, i.e., a subset of particles of the same species move farther than another subset at a given time period. Fig.\ref{dist} shows that particles that moved less than 0.8 \AA \: are well distributed in the samples. Between 0.9 and 1.2 \AA \: we found a plateau-like distribution of roughly one particle. For larger displacements, distribution goes down to zero as expected. The suggested picture is that a small but important fraction of SiO$_4$ tetrahedra performs jumps that make up for the viscous flow observed from the diffusion coefficient. The remaining particles move, on average, a small fraction of the bond length. These sluggish rearrangements end up creating room for the jump mediated diffusion to occur.

Additionally, Fig.\ref{dist} provides a concise analysis about caging and hopping processes. The well-distributed displacements can be associated with atomic vibrations occuring in spheric regions of radius close to 0.2 \AA \:. Note that for the two largest time intervals, the distributions are quite similar. On the other hand, the distribution height in the plateau changes considerably with the time interval applied. That reinforces the fact that diffusion is due to particles that perform jumps of sizes close to the Si-O bond length.

\section{Conclusions}

In this work, we tackled the problem of defining effective diffusion coefficients from alternative dynamical properties. We applied molecular dynamics simulations in the entire work. The interaction potential to model our simulated lithium disilicate melt has been reported to reproduce quite well structural properties of both liquid and glassy states. This work reports that dynamical properties are also well reproduced for melts between 900 and 2400K. Given the reliability of the potential, we were able to perform an analysis of the diffusion mechanisms on a realistic model for lithium disilicate. Results show that there is a decoupling between the structural relaxation time and the experimental viscosity. Also, the Eyring relation fails when viscosity is considered as a proxy of the strucutral relaxation. Direct comparison between the diffusion obtained in simulations and the experimental viscosity revealed the Eyring relation is valid on a strict sense at least down to 1400K. Finally, an analysis of the mobility of slow-diffusing Si-atoms showed two very distinct moving patterns. That observation confirmed the emergence of dynamical heterogeneities on the system in the undercooled regime. Our conclusion is that the much discussed dynamical heterogeneities do not lead necessarily to a breakdown of the Eyring relation. Rather, a decoupling between experimental viscosity and structural relaxation time obtained via MD takes place. Our study suggests that relaxation mechanisms are still not fully accounted for in MD simulations. For now, the Eyring relation continues to be a reliable measure of kinetic processes in oxide melts except near the glass transition.

\begin{acknowledgments}
The authors would like to thank the Coordena\c{c}\~{a}o de Aperfei\c{c}oamento de Pessoal de N\'{\i}vel superior (CAPES), the Conselho Nacional de
Desenvolvimento Cient\'{\i}fico e Tecnol\'{o}gico (CNPq), and the Funda\c{c}\~{a}o de Amparo \`{a} Pesquisa do Estado de S\~{a}o Paulo (FAPESP) for
financial support. We would also like to thank Prof. E. D. Zanotto for his insightful discussions.
\end{acknowledgments}


\bibliography{artigo_LS2}

\end{document}